\documentstyle[12pt]{article}

\begin{document}

\newcommand {\ops}        {{\it o}\rm{-Ps}}

\def\half{\textstyle{1\over2}\displaystyle}
\flushbottom

\begin{titlepage}

\begin{tabbing}
\` UT-ICEPP 95-01\\
\` March 1995\\
\end{tabbing}

\bigskip

\renewcommand{\thefootnote}{\fnsymbol{footnote}}
\begin{center}{\LARGE\bf
A search for massive neutral bosons\\
\vspace{5 mm}
in orthopositronium decay\footnote[1]
{Submitted for publication in Physics Letters B}
}\end{center}

\bigskip
\bigskip
\bigskip

\begin{center}
{\large T. Maeno\footnote[2]
{E-mail address: maeno@uticeaix1.icepp.s.u-tokyo.ac.jp},
M. Fujikawa, J. Kataoka, Y. Nishihara, \\
\vspace{2.5mm}
S. Orito, K. Shigekuni and Y. Watanabe}\end{center}
\bigskip
\begin{center}{\large Department of Physics, Faculty of Science,}\end{center}
\begin{center}{\large University of Tokyo, Tokyo 113, Japan}\end{center}
\renewcommand{\thefootnote}{\arabic{footnote}}
\setcounter{footnote}{0}

\bigskip
\bigskip

\begin{abstract}
We have searched for an exotic decay of orthopositronium into
a single photon and a short-lived neutral boson in the hitherto
unexplored mass region above 900 ${\rm keV}/{\it c}^{2}$, by
noting that this decay is one of few remaining candidates
which could explain the discrepancy of the orthopositronium decay-rate.
A high-resolution measurement of the associated photon energy spectrum
was carried out with a germanium detector to search for a sharp peak
from this two-body decay. Our negative result provides
the upper-limits of\mbox{ }$2.0 \times 10^{-4}$ on the branching
ratio of such a decay in the mass region
from 847 to 1013 ${\rm keV}/{\it c}^{2}$, and excludes the possibility
of this decay mode explaining the discrepancy
in the orthopositronium decay-rate.
\end{abstract}

\bigskip
\bigskip

\begin{center}{\large ICEPP}\end{center}
\begin{center}
{\large International Center for Elementary Particle Physics,}
\end{center}
\begin{center}{\large Faculty of Science, University of Tokyo}\end{center}
\begin{center}{\large 7-3-1 Hongo, Bunkyo-ku, Tokyo 113, Japan}\end{center}

\bigskip

\end{titlepage}
\setcounter{page}{2}

Over the last seven years,
three precision measurements of the orthopositronium ($\ops$) decay-rate
were performed, which report
the decay-rate {\it faster} than the QED prediction \cite{qed}
by 1400, 1300 and 920 ppm, respectively,
corresponding to 6.7, 6.3 and 4.3 standard deviations \cite{newlife}.
In attempts to explain the origin of this discrepancy,
various exotic $ \ops $ decay modes
have been investigated without obtaining any evidence so far.
The decay mode into invisible final
states was ruled out at 3 ppm \cite{invi}.
The decay into two and four photons, forbidden by space-rotational
invariance and by the charge-conjugation invariance of QED,
was excluded at 233 and 8 ppm, respectively \cite {cvio,4r}.
The decay into two photons and a long-lived, weakly-interacting vector boson
was excluded at 10 ppm \cite {3r}.
The decay into a photon and a long-lived, weakly-interacting particle
was also ruled out at 10 ppm
for a mass less than 1010 ${\rm keV}/{\it c}^{2}$ \cite{rx3},
while that into a photon and a short-lived neutral boson
with a mass less than 900 $ {\rm keV}/{\it c}^{2} $
was excluded at 400 ppm \cite {slnbh,slnbl,slnba}.
One of very few remaining candidates for the source of the discrepancy
is the decay into a single photon
and a short-lived neutral boson $X ^{0}$ in
the hitherto unexplored mass range
above 900 ${\rm keV}/{\it c}^{2}$.
We report in this letter a dedicated search for such a decay.

A schematic view of the experimental setup is shown in Fig. 1.
$\rm{A} \  ^{22}$Na positron source of intensity 0.7 $ \mu \rm{Ci} $
(2 mm spot diameter) is
sandwiched between a plastic scintillator (NE104) of
thickness 100 $ \mu \rm{m} $
and a mylar sheet of thickness 25 $ \mu \rm{m} $,
each with a diameter of 12 mm.
The source and a target of silica aerogel
(dimensions $ 10 \times 10 \times 5 \, \rm{mm}^{3}$,
density $0.1 \, \rm{g \, cm}^{-3}$)
are held at the center of a pipe (18 mm diameter)
made of 20$ \mu \rm{m} $ thick aluminized mylar.
The pipe is filled with $\rm{N}_{2}$ gas at 1 atmosphere.
Most of the positrons emitted toward the target pass through
the scintillator, giving light pulses to
two photomultipliers (Hamamatsu H-3165-PV).
When stopped in the silica aerogel,
they form positronium-atoms for about 20$\%$ of cases.
The total efficiency of $ \ops $ formation
in our set-up was measured to be 3.0$\%$
per $\beta ^{+} \rm{decay}$.

The energy of the photon from the decay of $ \ops $ is measured
by a planar high-purity germanium detector
(diameter 16 mm, thickness 10 mm, Ortec GLP-16195/10-P).
The germanium detector has a thin Be window of thickness 0.13 mm,
which results to a high efficiency to low-energy photons down to 5 keV.
The energy resolution and
the absolute peak efficiencies of the germanium detector
are determined as a function of the photon-energy
by using the line $ \gamma $
peaks from various sources of
known strength, i.e., $ ^{22}\rm{Na},\ ^{57}\rm{Co},
\ ^{133}\rm{Ba},
\ ^{152}\rm{Eu}, \ ^{210}\rm{Pb} \ \rm{and} \ ^{226}\rm{Ra}, $
placed at the source position.
The energy resolution obtained is 270,
525 and 828 eV FWHM at 14.4 , 136.5 and 356.0 keV,
respectively.

The trigger and the data-acquisition system is arranged as follows.
The pulses from the two photomultipliers
are discriminated individually and the coincidences between
them provide the start
signals to the time-to-digital converter (TDC ; LeCroy 2228A),
as well as the
trigger signals to the CAMAC system. One output from the preamplifier of
the germanium detector is fed through
a fast-filter amplifier (Ortec 474) with an integration-time of 100 nsec,
into the peak-hold analog-to-digital converter (ADC-1 ; Hohshin C008),
and also into a discriminator
whose output is used as the stop signal for the TDC.
The other preamplifier output is amplified by a shaping amplifier (Ortec 572)
with an integration-time of 3 $ \mu \rm{sec} $.
One output of the shaping amplifier is
fed into the peak-hold analog-to-digital converter
(ADC-2 ; Hohshin C011),
to record the information of the low-energy region below 160 keV.
The other output of the shaping amplifier is attenuated and fed
into the other input of the ADC-2,
to record the energy information in a wider range up to 600 keV.
If the TDC stop signal is not generated
within 5 $ \mu \rm{sec} $ after the trigger,
the data-readout is terminated and the system is
cleared to accept the next trigger.

The data were collected from 10 runs
for a total of 2.26 $ \times \ 10^{6} \ \rm{sec} $.
Throughout the data-taking period, the room temperature
was controlled to $ \pm 0.4 ^{\circ} \rm{C} $.
The energy calibrations and the measurements of the energy
resolution were carried out
at intervals of 7 days.
In addition, the position and the width of the 511 keV annihilation
line were monitored
by offline analysis of the data, which
showed a stable peak position within 0.03 keV
and a constant energy-resolution throughout the data-taking period.

The offline selection and calibration process is as follows.
In order to eliminate the pile-up effects,
we demand the consistency
between two ADC values with different integration-time :
If the ratio of the two values deviates
from the central value by more than 6$\%$,
the event is rejected as being affected by the pile-up.
This selection has an efficiency of 92$\%$.
The time walk of the discriminator output is corrected
by using the pulse height information from the ADC-2.
The nonlinearlity of the TDC is corrected
by using a flat spectrum between random triggers and clock signals.

Figure 2 shows the time spectrum between the scintillator and
germanium signals.
A sharp peak of the prompt annihilations is followed by the exponential
decay of $ \ops $ and subsequently by the constant accidentals.
The time spectrum fits well to an exponential function plus a constant term.
A decay time of 134.2 $\pm$ 0.4 nsec is consistently obtained for
various time spans
if the fitting starts more than 150 nsec after the prompt peak.
To obtain a pure sample of the $ \ops $ decay, we select the events
in the time-window between 150 and 400 nsec after the prompt annihilations.
The accidental contribution is measured by using the events
in the time-window between 2000 and 3200 nsec after the prompt peak.

Figures 3a and 3b show the measured energy-spectra from the $ \ops $ decay
($ \ops $ spectrum), thus selected
and the accidental contribution being subtracted.
The two-body decay into a photon and $ X ^{0} $ ($ \gamma X ^{0} $ decay)
would appear in this $ \ops $ spectrum
as a narrow peak on a smooth background.
The peak position $k_{p}$ is related to the $X^{0}$ mass, $m_{X^{0}}$, by
\begin{displaymath}
k_{p} = m_{e} \left[1-\left(m_{X^{0}}/2m_{e} \right)^{2} \right] \, ,
\end{displaymath}
where $m_{e}$ is the electron mass.

Peaks in the $ \ops $ spectrum
have been searched for
by scanning $k_{p}$ with a step of 0.088 keV from 10 to 160 keV,
corresponding to the $ m_{X^{0}}$ region
from 847 to 1013 $ \rm{keV}/{\it c}^{2} $ :
At each $k_{p}$, the $ \ops $ spectrum was fitted with the function
\begin{displaymath}
S(k) + \ {\bf n}(k_{p}) P(k) \; ,
\end{displaymath}
where $ S(k) $ and $ P(k) $ represent
the smooth background-spectrum and
the normalized peak-function, respectively.
The fitting parameter ${\bf  n}(k_{p})$ corresponds to the number of photons
from the $ \gamma X^{0} $ decay events under the peak at $k_{p}$.
The background-spectrum $ S(k) $ is obtained at each $k_{p}$
by fitting the $ \ops $ spectrum
(excluding the region within 2.5 times the FWHM of the peak position)
with a polynomial function of up to seventh order.
Since the intrinsic width of $X^{0}$ is expected to be extremely small,
the observable width of the peak would
be dominated by the energy-resolution of the detector
and by the thermal motion of $\ops$.
The thermal kinetic energy of $\ops$ has been
measured in an aerogel similar to the one used in this experiment
and is known to decrease to
about 0.04 eV within 60 nsec after formation \cite{aero,powd}.
The Doppler-broadening due to this thermal kinetic energy
would contribute negligibly
to the peak-width.
Therefore the shape of the peak is assumed to be Gaussian,
of a width determined by the energy-resolution of the germanium detector.

No statistically significant peak beyond 4 standard deviations was detected
in the $\ops$ spectrum, resulting to upper limits on ${\bf n}(k_{p})$.
The limits on ${\bf n}(k_{p})$ can be converted into the corresponding limits
on the branching ratio of the $\gamma X^{0}$ decay,
$Br(k_{p})$, by using the formula
\begin{displaymath}
Br(k_{p}) = \frac{{\bf n}(k_{p})}{\epsilon(k_{p}) \, N_{ops}} \; ,
\end{displaymath}
where $\epsilon(k_{p})$ is the absolute
peak efficiency for a $\gamma$-ray of energy $k_{p}$.

$N_{ops}$ represents the total number of the $ \ops $ decays
occurring within the 150 - 400 nsec time-window,
and is determined by
fitting the ``compton-free'' region of the $ \ops $ spectrum
(between 400 to 490 keV, where the three-photon decays dominate)
with the theoretical spectrum \cite{adkins} of the three-photon decay
folded with the energy-dependent efficiency.
It should be noted that $Br(k_{p})$ thus
obtained has a small systematic error, being
dependent only on the relative values of
the efficiencies, not on their absolute values.

The resulting upper-limits at a $90\%$ confidence level
on the branching ratio of $ \gamma X ^{0} $ decay are shown in Fig. 4 ,
as a function of $ m_{X ^{0}} $.
As indicated in this figure, new upper limits on the $ \gamma X ^{0} $
branching ratio in the mass range of $ X ^{0} $ from 900 to
1013 $ \rm{keV}/{\it c}^{2} $ have been obtained.
Since our upper limits are at least a factor of 7 below the
level which would explain the reported discrepancy of the $ \ops $ lifetime,
we conclude that the $ \ops $ decay mode of $ \gamma X ^{0} $ in
the mass region
below 1013 $ \rm{keV}/{\it c}^{2} $
can not explain the reported discrepancy of the orthopositronium decay-rate.

We wish to thank Dr. S. Asai for suggestions and discussions,
Dr. X. Zhang for kindly lending us the germanium detector,
Mr. Y. Nakamura, Mr. N. Kimura and Mr. M. Matsui,
of Japan Radioisotope Association, for preparing the positron source.
The analysis was performed on the RS/6000 workstations
supplied for partnership program between ICEPP and IBM Japan Ltd.

\newpage

\bigskip
\bigskip
\bigskip

\begin{center}{\Large \bf Figure captions}\end{center}

\bigskip

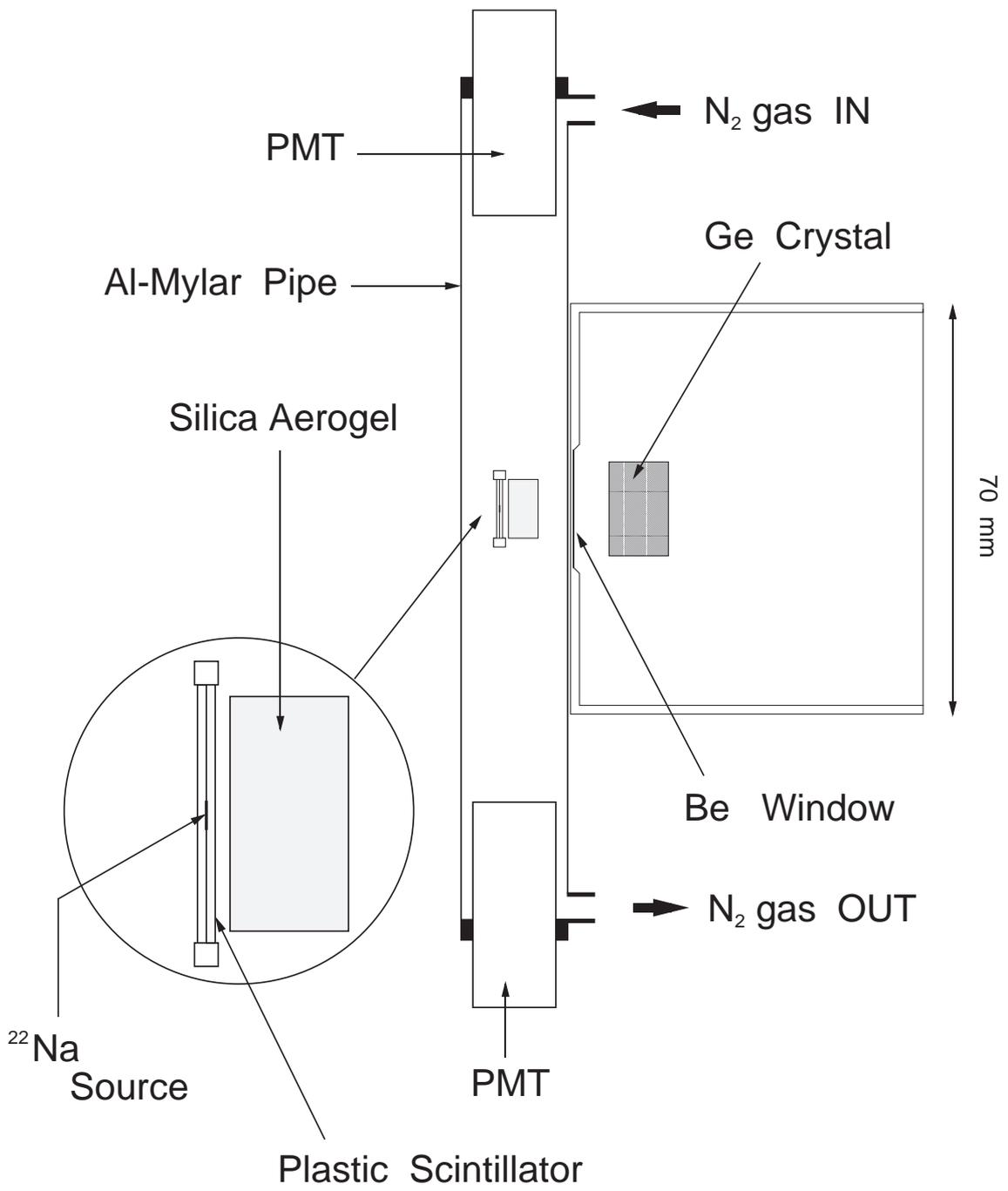
Fig. 1. A schematic view of
the experimental setup. Circular inset: a magnified view
of the source region.

\bigskip

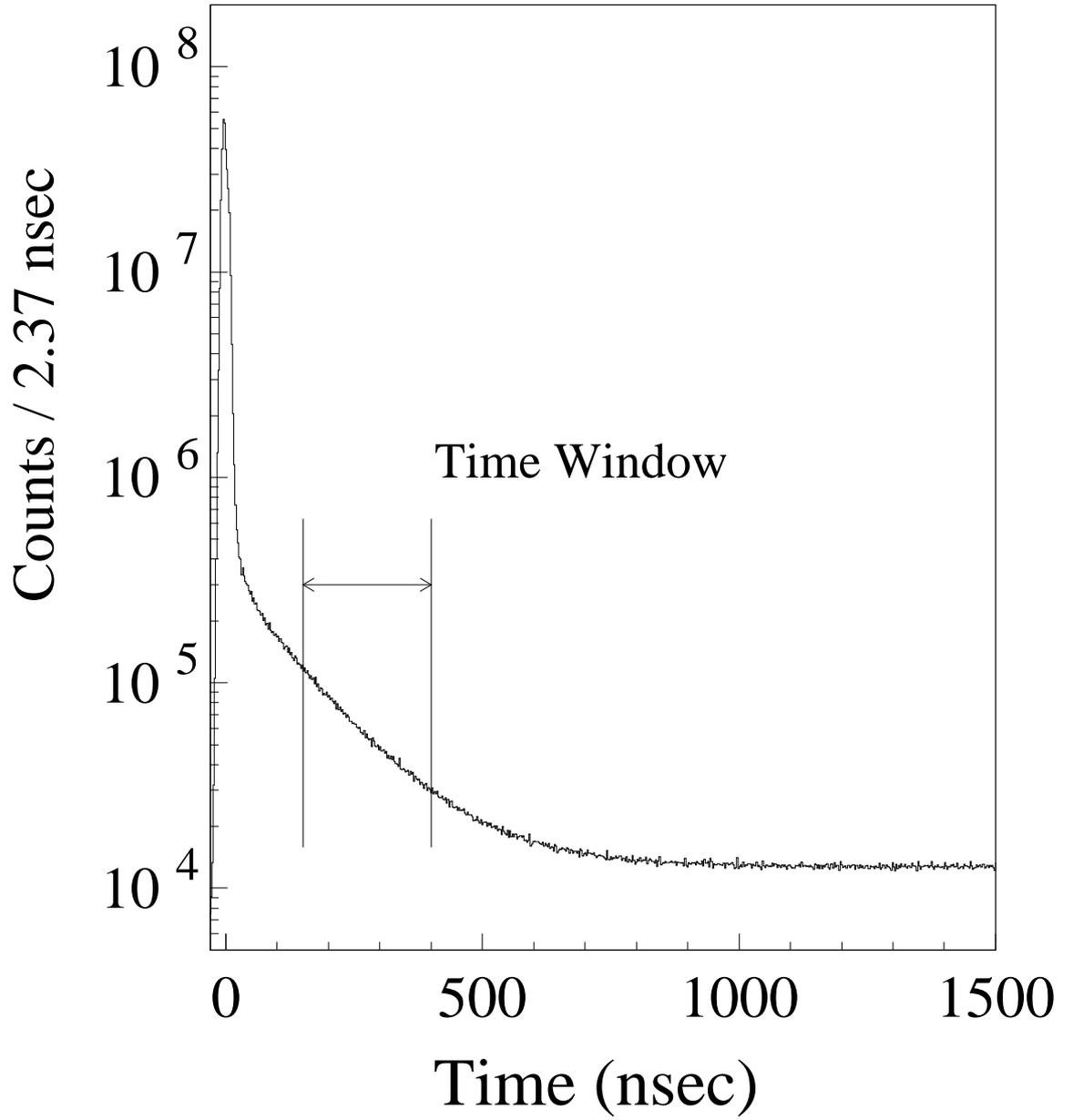
Fig. 2. The time spectrum between the scintillator and
germanium detector signals.

\bigskip

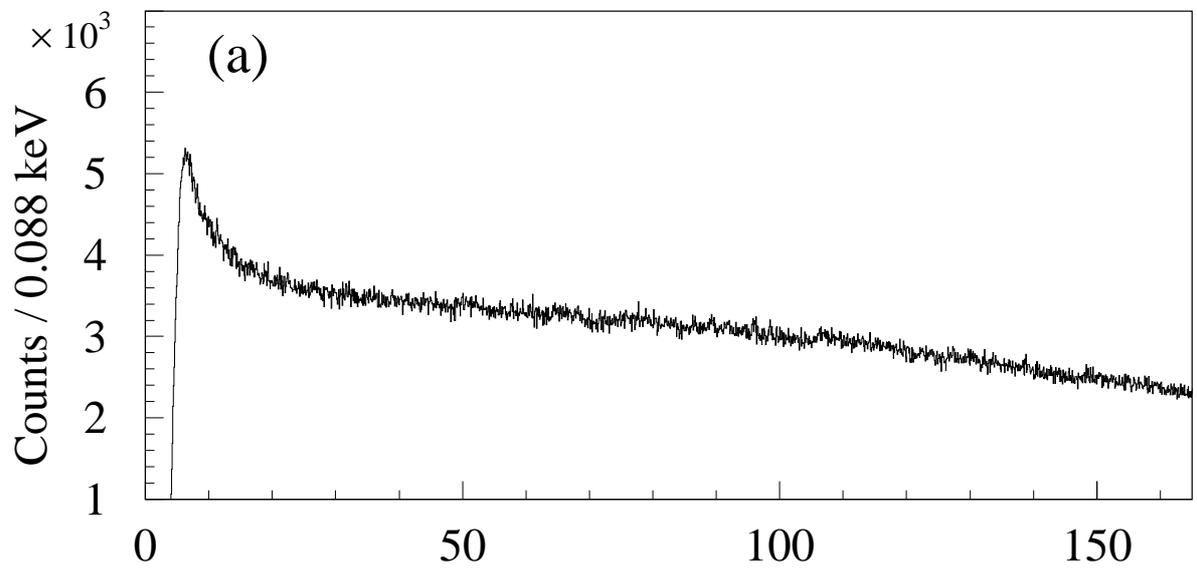
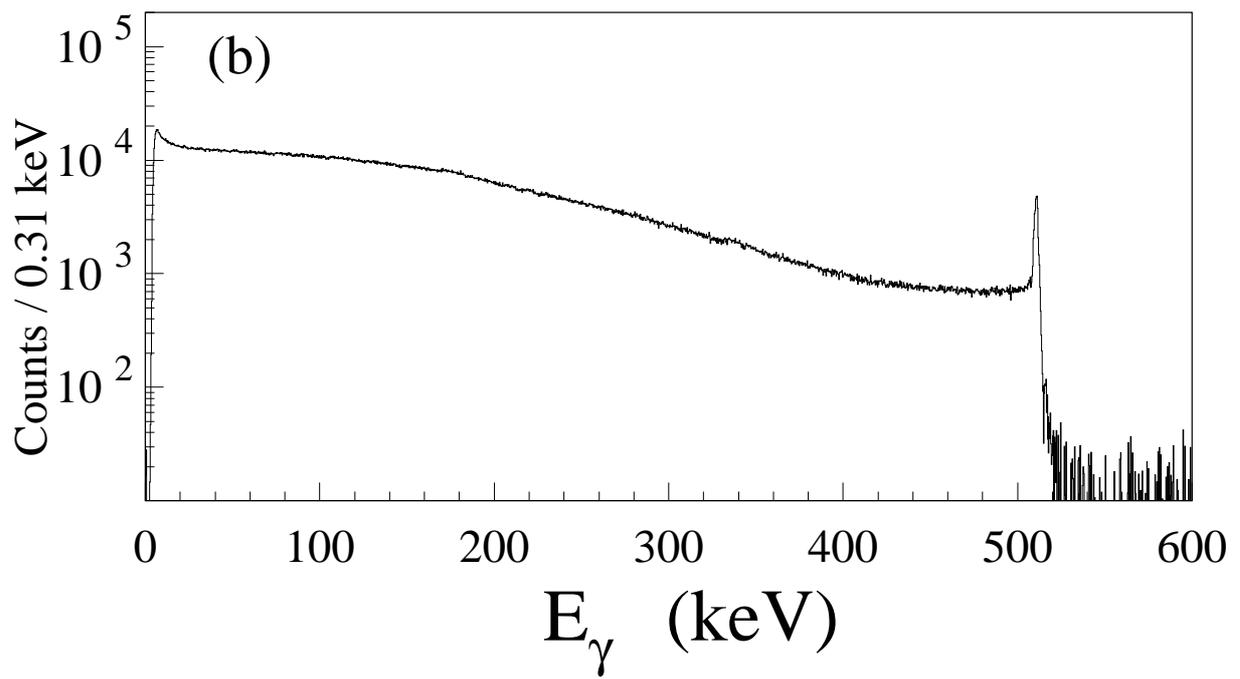
Fig. 3. The $ \ops $ spectra after subtracting the
accidental contribution. (a) Below 160 keV. (b) Up to 600 keV.

\bigskip

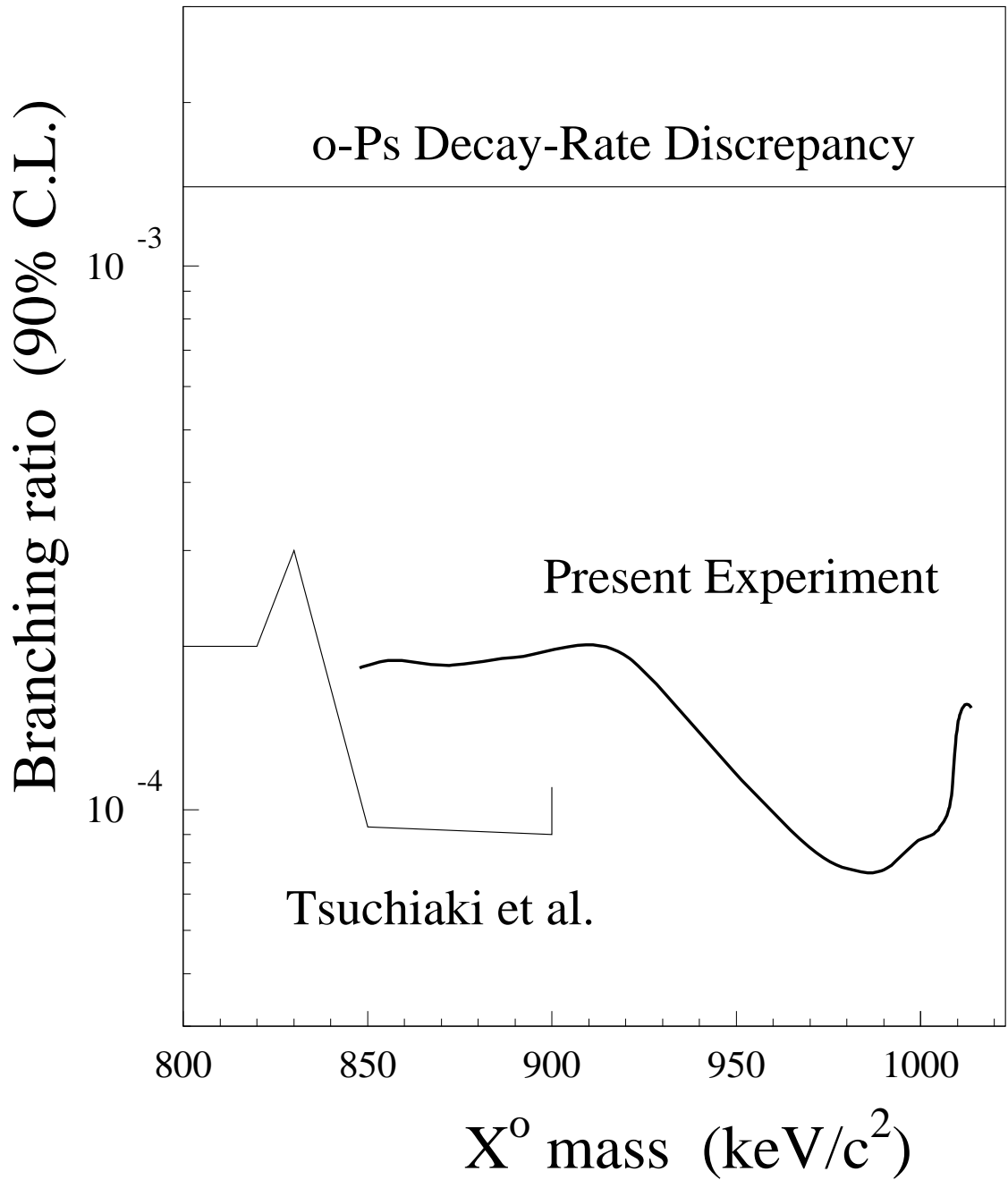
Fig. 4. The resultant upper-limits at 90\% C.L. on the
branching ratio of $ \gamma X ^{0} $ decay in comparison with the
existing limits \cite{slnbh}.

\end{document}